\documentclass[11pt]{article}
\usepackage[sc]{mathpazo} %Like Palatino with extensive math support
\usepackage{fullpage}
\usepackage[authoryear,sectionbib]{natbib}
\usepackage[utf8]{inputenc}
\usepackage{titlesec}

\titleformat{\section}[block]{\Large\bfseries\filcenter}{\thesection}{1em}{}
\titleformat{\subsection}[block]{\Large\itshape\filcenter}{\thesubsection}{1em}{}
\titleformat{\subsubsection}[block]{\large\itshape}{\thesubsubsection}{1em}{}
\titleformat{\paragraph}[runin]{\itshape}{\theparagraph}{1em}{}[. ]

\usepackage{graphicx,amsmath,amssymb}

\title{Niche theory for mutualism:\\ A graphical approach to plant-pollinator network dynamics}
\author{Fernanda S. Valdovinos$^{1,\ast}$\\
	Robert Marsland III$^{2}$}

\date{\today}

\begin{document}
	
	\maketitle
	
	\noindent{} 1. University of California - Davis, Davis, CA 95616;
	
	\noindent{} 2. Boston University, Boston, Massachusetts 02215;
	
	\noindent{} $\ast$ Corresponding author; e-mail: fvaldovinos@ucdavis.edu.
	
	\bigskip
	
	\textit{The authors wish to be identified by the reviewers.}
	
	\bigskip
		
	
	\bigskip
	
	\textit{Number of words in the main text (excluding abstract, figure legends, tables, and literature cited)}: 5,810.
	
	\bigskip
	
	%An example, to be replaced by our actual figures:
	\textit{Manuscript elements}: Figure~1, figure~2, figure~3, figure~4, table~1, table~2, online appendices~A, B, C and D including figures~S1, S2, S3 and S4.
	
	\bigskip
	
	\textit{Manuscript type}: Article. %Or e-article, note, e-note, natural history miscellany, e-natural history miscellany, comment, reply, invited symposium, or countdown to 150.
	
	\bigskip
	
	\textit{Code producing figures available at}: github.com/robertvsiii/niche-mutualism
	
	\noindent{\footnotesize Prepared using the suggested \LaTeX{} template for \textit{Am.\ Nat.}}
	
	%\linenumbers{}
	%\modulolinenumbers[3]
	
	\newpage{}
	
	\section*{Abstract}
	
	Contemporary Niche Theory is a useful framework for understanding how organisms interact with each other and with their shared environment. Its graphical representation, popularized by Tilman's Resource Ratio Hypothesis, facilitates the analysis of the equilibrium structure of complex dynamical models including species coexistence. This theory has been applied primarily to resource competition since its early beginnings. Here, we integrate mutualism into niche theory by expanding Tilman's graphical representation to the analysis of consumer-resource dynamics of plant-pollinator networks. We graphically explain the qualitative phenomena previously found by numerical simulations, including the effects on community dynamics of nestedness, adaptive foraging, and pollinator invasions. Our graphical approach promotes the unification of niche and network theories, and deepens the synthesis of different types of interactions within a consumer-resource framework.
	
	\section*{Secondary Abstract}
	\textit{Teoría de Nicho para Mutualismos: Una aproximación gráfica a la dinámica de redes planta-polinizador}
	
	La Teoría Contemporánea de Nicho es un marco útil para entender cómo los organismos interactúan entre ellos y con su ambiente compartido. Su representación gráfica, popularizada por la Hipótesis de Razón de Recursos de Tilman, facilita el análisis de la estructura de equilibrio de modelos dinámicos complejos, incluyendo la coexistencia de especies. Esta teoría ha sido aplicada primariamente a competencia por recursos desde sus inicios. Aquí, integramos el mutualismo dentro de la teoría de nicho al expandir la representación gráfica de Tilman al análisis de la dinámica consumidor-recurso de las redes planta-polinizador. Explicamos gráficamente fenómenos cualitativos encontrados previamente mediante simulaciones numéricas, incluyendo los efectos sobre la dinámica comunitaria del anidamiento, forrajeo adaptativo y de las invasiones por polinizadores. Nuestra aproximación gráfica promueve la unificación de las teorías de nicho y de redes, y profundiza la síntesis de diferentes tipos de interacciones dentro de un marco de consumidor-recurso.
	
	\newpage{}

\section*{Introduction}
Mutualistic interactions pervade every type of ecosystem and level of organization on Earth (\citealt{boucher1982ecology,bronstein2015study}). Mutualisms such as pollination (\citealt{ollerton2011flowering}), seed dispersal (\citealt{wang2002closing}), coral symbioses (\citealt{rowan2004coral}), and nitrogen-fixing associations between plants and legumes, bacteria, or fungi (\citealt{horton2001molecular,heijden2008unseen}) sustain the productivity and biodiversity of most ecosystems on the planet and human food security (\citealt{potts2016safeguarding,ollerton2017pollinator}). However, ecological theory on mutualisms has been scarce and less integrated than for predation and competition, which hinders our ability to protect, manage, and restore mutualistic systems (\citealt{vandermeer1978varieties,bascompte2014mutualistic,bronstein2015study}). This scarce theoretical development is of particular concern because several mutualisms such as coral-algae and plant-pollinator that play a critical role in the functioning of ecosystems are currently under threat (\citealt{brown1997coral,rowan2004coral,goulson2015bee,ollerton2017pollinator}). In particular, Niche Theory (\citealt{MacArthur1969,macarthur1970species,tilman_resource_1982,leibold1995niche,chase2003ecological}) for mutualisms has only recently started to be developed (\citealt{peay2016mutualistic,johnson2019coexistence}). \cite{chase2003ecological} suggest that Contemporary Niche Theory can be expanded to mutualism, but such suggestion has yet to be explored. Here, we expand niche theory to mutualistic networks of plant-pollinator interactions by further developing the graphical approach popularized by \cite{tilman_resource_1982} to analyze a consumer-resource dynamic model of plant-pollinator networks developed, analyzed, and tested by \cite{valdovinos2013adaptive,valdovinos2016niche,valdovinos2018species}.

For about 70 years, theoretical research analyzing the population dynamics of mutualisms roughly only assumed Lotka-Volterra type models (sensu \citealt{valdovinos2019mutualistic}) to conduct their studies (e.g., \citealt{kostitzin1934parasitisme,gause1935behavior,vandermeer1978varieties,wolin1984models,bascompte2006asymmetric,okuyama2008network,bastolla2009architecture}). Those models represent mutualistic relationships as direct positive effects between species using a (linear or saturating) positive term in the growth equation of each mutualist that depends on the population size of its mutualistic partner.  While this research increased our understanding of the effects of facultative, obligate, linear, and saturating mutualisms on the long-term stability of mutualistic systems, more sophisticated understanding of their dynamics (e.g., transients) and of phenomena beyond the simplistic assumptions of the Lotka-Volterra type models was extremely scarce. A more mechanistic consumer-resource approach to mutualisms has been recently proposed by Holland and colleagues (\citealt{holland2005mutualisms,holland2010consumer}) and further developed by \cite{valdovinos2013adaptive,valdovinos2016niche,valdovinos2018species}. This approach decomposes the net effects assumed always positive by Lotka-Volterra type models into the biological mechanisms producing those effects including the gathering of resources and exchange of services.

The key advance of the consumer-resource model developed by \cite{valdovinos2013adaptive} is separating the dynamics of the plants’ vegetative biomass from the dynamics of the plants’ floral rewards. This separation allows: i) tracking the depletion of floral rewards by pollinator consumption, ii) evaluating exploitative competition among pollinator species consuming the floral rewards provided by the same plant species, and iii) incorporating the capability of pollinators (adaptive foraging) to behaviorally increase their foraging effort on the plant species in their diet with more floral rewards available. Another advance of this model is incorporating the dilution of conspecific pollen carried by pollinators, which allows tracking competition among plant species for the quality of pollinator visits (see next section).

This contribution analyzes the dynamics of plant-pollinator networks when all the above-mentioned biological mechanisms are considered. Specifically, we provide analytical understanding for the results found with extensive numerical simulations (\citealt{valdovinos2013adaptive,valdovinos2016niche, valdovinos2018species}, hereafter ``previous simulations"), and generalize some of them beyond the original simulation conditions. By ``analytical understanding" we refer to finding those results using a graphical approach whose geometry rigorously reflects mathematical analysis (\citealt{tilman_resource_1982,koffel2016geometrical}, also provided in our Appendices). Our Methods describe the Valdovinos et al.'s model and our graphical approach, including the conditions for coexistence among adaptive pollinators sharing floral rewards and how we use projections to analyze high-dimensional systems. Our Results first demonstrate the effects of nestedness on species coexistence in networks without adaptive foraging found by previous simulations (\citealt{valdovinos2016niche}). Nestedness is the tendency of generalists (species with many interactions) to interact with both generalists and specialists (species with one or a few interactions), and of specialists to interact with only generalists. Second, we demonstrate the effects of adaptive foraging on species coexistence in nested networks found by the same simulation study. Third, we demonstrate the impacts of pollinator invasions on native pollinators in  nested networks with adaptive foraging found numerically by \cite{valdovinos2018species}. Finally, we discuss how our approach helps to integrate niche and network theories, and deepens the synthesis of different types of interactions within a consumer-resource framework.

\section*{Methods}

\subsection*{1. Dynamical model of plant-pollinator interactions}

\cite{valdovinos2013adaptive} model the population dynamics of each plant and pollinator species of the network, as well as the dynamics of floral rewards 
and pollinators' foraging preferences (see Table 1 for definitions of variables and parameters). Four functions define these dynamics. The function $V_{ij}(p_i,a_j) =  \alpha_{ij}\tau_{ij} a_j p_i$ represents the visitation rate of animal species $j$ to plant species $i$ and connects the dynamics of plants, animals, rewards, and foraging preferences. An increase in visits increases plant growth rate via pollination and animal growth rate via rewards consumption, but decreases rewards availability. The function $\sigma_{ij}(p_k) = \frac{\alpha_{ij} \tau_{ij} p_i}{\sum_{k\in P_j} \alpha_{kj}\tau_{kj} p_k}$ represents the fraction of $j$’s visits that successfully pollinate plant $i$, and accounts for the dilution of plant $i$’s pollen when $j$ visits other plant species. Pollinators visiting many different plant species carry more diluted pollen (low quality visits) than the pollen carried by pollinators visiting only one plant species (high quality visits). The function $\gamma_i(p_k) = g_i (1-\sum_{l\neq i \in P} u_l p_l - w_i p_i)$ represents the germination rate of the seeds produced by the successful pollination events, where $g_i$ is the maximum fraction of $i$-recruits subjected to both inter-specific ($u_l$) and intra-specific ($w_i$) competition. Finally, the function $f_{ij}(R_i/p_i) = b_{ij}\frac{R_i}{p_i}$ represents the rewards consumption by animal $j$ in each of its visits to plant $i$. These functions capturing the above mentioned biological processes lead to the following equations:

\begin{align}
\frac{dp_i}{dt} &= {\overbrace{\gamma_i(p_k)}^{\rm germination\, rate}} \sum_{j \in A_i} {\overbrace{e_{ij}\sigma_{ij}(p_k) V_{ij}(p_i,a_j)}^{\rm seed\, production}} -{\overbrace{\mu_i^P p_i}^{\rm mortality}}\label{eq:p}\\
\frac{da_j}{dt} &= \sum_{i\in P_j} c_{ij}{\overbrace {V_{ij}(p_i,a_j) f_{ij}(R_i/p_i)}^{\rm rewards\, consumption}} - {\overbrace{\mu_j^A a_j}^{\rm mortality}} \label{eq:a}\\
\frac{dR_i}{dt} &= p_i{\overbrace{\left[\beta_i  - \phi_i \frac{R_i}{p_i}\right]}^{\rm per-plant\, rewards\, production}} - \sum_{j\in A_i} {\overbrace{V_{ij}(p_i,a_j) f_{ij}(R_i/p_i)}^{\rm rewards\, consumption}}.\label{eq:R}\\
\frac{d \alpha_{ij}}{dt} &= \frac{G_j \alpha_{ij}}{a_j} \left(c_{ij} {\overbrace{V_{ij}^s(p_i,a_j)f_{ij}(R_i/p_i)}^{\rm rewards\,consumption\,as\,specialist}}- \sum_{k \in P_j} c_{kj} {\overbrace{ V_{kj}(p_k,a_j)f_{kj}(R_k/p_k)}^{\rm actual\, rewards\, consumption}}\right),\label{eq:adapt}
\end{align}
where $V_{ij}^s = \tau_{ij}a_j p_i$ is the visitation rate of animal species $j$ to plant species $i$ under a pure specialist strategy $\alpha_{ij} = 1$. That is, the preference of animal $j$ for plant $i$ increases when the rewards that could be extracted from plant species $i$ by application of full foraging effort to that plant ($\alpha_{ij} = 1$) exceed the rewards currently obtained from all plants in $j$’s diet. The preference decreases in the opposite case, where the rewards obtainable by exclusive foraging on plant $i$ are lower than the current rewards uptake level. Note that the terms in Eq. (\ref{eq:adapt}) have been re-arranged from previous publications of this model to emphasize the coupling of the four equations through the visitation rates $V_{ij}$. We use parentheses that include the variables determining each of the functions in the equations to distinguish functions from parameters, but in the text those parentheses are excluded for better readability. The visitation rate $V_{ij}$ and the rewards extracted per visit $f_{ij}$ can also be modeled by a saturating function following Holling's Type II functional response (\citealt{holling1959some}), as discussed in Appendix D.  
\begin{table}
	\begin{tabular}{c|l}
		Symbol & Meaning\\
		\hline
		$p_i$ & plant abundance per unit area ([ind.]/m$^2$)\\ 
		$a_j$ & animal (pollinator) abundance per unit area ([ind.]/m$^2$)\\
		$R_i$ & reward abundance per unit area (g/m$^2$)\\
		$\alpha_{ij}$ & foraging preference (dimensionless)\\ 
		\hline
		$g_i$ & max germination rate ([ind.]/[seeds])\\
		$u_l$ & plant inter-specific competition (m$^2$/[ind.])\\
		$w_i$ & plant intra-specific competition (m$^2$/[ind.])\\
		$e_{ij}$ & expected number of seeds per pollination event ([seeds]/[visits])\\
		$\tau_{ij}$ & visitation efficiency ([visits]m$^2$/[ind.]$^2$ yr)\\
		$\mu_i^{P/A}$ & mortality rates (1/yr)\\
		$c_{ij}$ & conversion efficiency of rewards into animal abundance ([ind.]/g)\\
		$b_{ij}$ & per-visit rewards extraction (1/[visits])\\
		$\beta_i$ & per-plant reward production (g/[ind.]yr)\\
		$\phi_i$ & self-limitation of reward production (1/yr)\\
		$G_j$ & adaptation rate (dimensionless)\\
		\hline
	\end{tabular}
	\caption{Model variables and parameters}
	\label{tab:par}
\end{table}

The sums in equations (1-4) are taken over the sets of $A_i$ and $P_j$ of pollinator species that are capable of visiting plant $i$, and plant species that can be visited by pollinator $j$, respectively. Those sets are defined by the network structure taken as model input. Finally, the dynamic preferences of Eq. (\ref{eq:adapt}) model adaptive foraging. These preferences are restricted by $\sum_{k \in P_j} \alpha_{kj} = 1$. When adaptive foraging is not considered, foraging preferences are fixed to:

\begin{align}
\alpha_{ij} = \frac{1}{P_j}\label{eq:fixed}
\end{align}
where $P_j$ here represents the number of plant species visited by pollinator species $j$.

\subsection*{2. Niche theory for plant-pollinator dynamics}

\begin{figure*}
	\includegraphics[width=16cm]{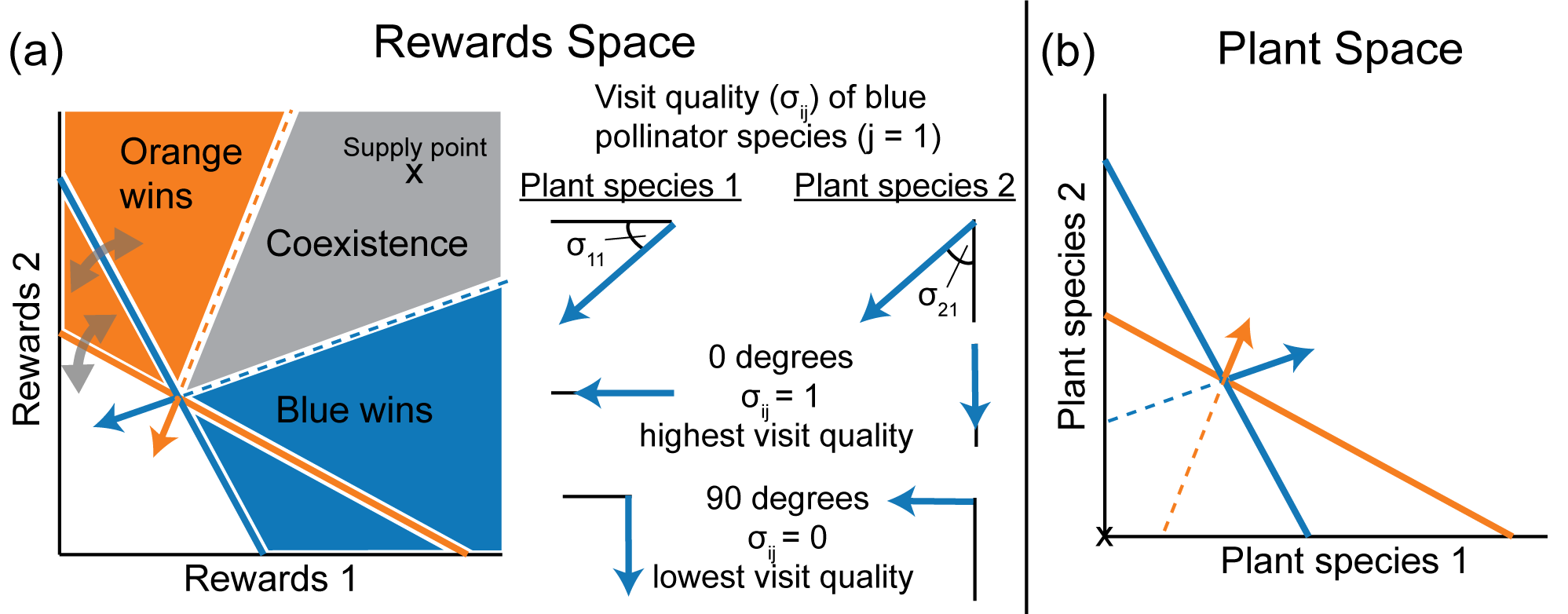}
	\caption{\linespread{1.3}\selectfont{} {\bf Niche theory for mutualism.} \emph{(a)} Representation of plant-pollinator system as a standard consumer-resource type model, for timescales on which plant populations are approximately constant. Impact vectors and ZNGIs are shown for two pollinator species (blue and orange) competing for the rewards of two plant species. Adaptive foraging causes the ZNGIs and impact vectors to rotate in the direction of the most abundant resource, as discussed in detail in Appendix A. The angle between the impact vector and a given rewards axis affects the pollinator's visit quality for the corresponding plant, with zero degrees corresponding to $\sigma_{ij}=1$ (highest visit quality), and 90 degrees corresponding to $\sigma_{ij} = 0$ (lowest visit quality). See supplementary figure \ref{fig:angle} for detailed discussion of angle-quality relationship. \emph{(b)} Representation in terms of plant populations for analysis of longer timescales, where the mutualism becomes visible. The ``supply point'' is now located at the origin, and the pollinator impacts are necessary to sustain nonzero plant abundance. The location of the ZNGIs depends on the current nutritional value of each plant species, which is lower for species whose floral rewards are more depleted. The impact vectors (see Table \ref{tab:map}) depend on both the visit quality and the per-capita visit frequency of each pollinator species ($\sigma_{ij}$ and $V_{ij}/a_{j}$ of Eq. (\ref{eq:p}), respectively), and encode each pollinator's contribution to the total number of seedlings in the next generation.}
	\label{fig:reward-plant}
\end{figure*}

``Niche" is a central concept in ecology, significantly clarified and refined over the past fifty years (\citealt{MacArthur1969,macarthur1970species,tilman_resource_1982,leibold1995niche,chase2003ecological}). Here, we analyze the niche of plant and pollinator species within their mutualistic interactions, assuming all their other niche variables (e.g., soil nutrients, water, temperature, nesting sites) constant and sufficient for supporting their populations. There are two reasonable choices for the definition of environment space in plant-pollinator systems. First, on short timescales (i.e., within a flowering season, Fig.~\ref{fig:reward-plant}\emph{a}), the plant populations can be regarded as constant and the relevant environmental factors are the floral rewards. Second, on longer timescales (i.e., across several flowering seasons, Fig.~\ref{fig:reward-plant}\emph{b}), plant populations represent the axes for the environment space, letting the reward levels implicitly determine the value of each plant population as a food source. Table \ref{tab:map} summarizes both representations in terms of the model parameters. This section explains both representations to provide a broader picture of niche theory applied to plant-pollinator systems, but we obtain our results on Rewards Space.

The ``requirement niche'' of each pollinator species $j$ ($j = 1, 2 \dots A$) in either Rewards or Plant Space can be encoded by a zero-net-growth isocline (ZNGI) (\citealt{tilman_resource_1982,leibold1995niche}). The ZNGI is a hypersurface that separates the environmental states where the growth rate is positive from the states where it is negative. Environmental states along the ZNGI support animal reproduction rates that exactly balance mortality rates, leading to constant population sizes. Adaptive foraging allows the ZNGIs in Rewards Space to dynamically rotate in the direction of the most abundant rewards. The ZNGIs are dynamic in Plant Space (even in the absence of adaptive foraging) because the contribution each plant makes to the animal growth rate depends on the current reward level.

The ``impact niche'' of each pollinator species is represented by an \emph{impact vector}, which specifies the magnitude and direction of the environmental change induced by an average individual of the species (\citealt{tilman_resource_1982,leibold1995niche}).  In Rewards Space, the impact of a pollinator species is the rate at which it depletes the floral rewards, just as in traditional models of resource competition, but its angle takes on a new importance in connection with the visit quality $\sigma_{ij}$. A nearly perpendicular impact vector to a given rewards axis means that only a small fraction of the pollinator's visits are allocated to the corresponding plant, and most of the pollen carried by this pollinator belongs to other plant species. A plant species will eventually go extinct if all its visits have such low quality (see below). Note that the exact mapping from the angle to the visit quality depends on the foraging strategy, number of plant species, and plant abundances (see Fig. \ref{fig:angle} of Appendix C). In Plant Space, the positive effects of plant-pollinator mutualisms are directly visible in the impact vectors pointing to larger plant population sizes (as opposed to pointing to smaller population sizes in the traditional models of resource competition), and represent the number of successful pollination events caused by each pollinator species.
\begin{table}
\centering
	\begin{tabular}{|c|c|c|}
		\hline
		\multicolumn{3}{|c|}{\bf Rewards Space}\\
		\hline
		{\bf Niche concept} & {\bf Description} & {\bf Mathematical expression} \\
		\hline
		ZNGI & Reproduction/mortality balance & $\sum_{i\in P_j} c_{ij} (V_{ij}/a_j) f_{ij} = \mu_j^A$ \\
		\hline
		Impact Vector & Per-capita rewards consumption & $-(V_{ij}/a_j) f_{ij}$ \\
		\hline
		Supply Point & Rewards equilibrium without animals & $\beta_i p_i/\phi_i$ \\
		\hline
		\hline
		\multicolumn{3}{|c|}{\bf Plant Space}\\
		\hline
		{\bf Niche concept} & {\bf Description} & {\bf Mathematical expression} \\
		\hline
		ZNGI & Reproduction/mortality balance & $\sum_{i\in P_j} c_{ij} (V_{ij}/a_j) f_{ij} = \mu_j^A$\\
		\hline
		Impact Vector & Plant production & $\gamma_i e_{ij} \sigma_{ij} (V_{ij}/a_j)$\\
		\hline
		``Supply Point'' & Plant equilibrium without animals & $0$\\
		\hline
	\end{tabular}
\caption{Mapping elements of the model to niche theory concepts.}
\label{tab:map}
\end{table}

The environment also has its intrinsic dynamics, represented by a \emph{supply vector} (\citealt{tilman_resource_1982,chase2003ecological}). In Rewards Space, the supply vector points towards the \emph{supply point} where the rewards reach equilibrium in the absence of pollinators (like in traditional models of resource competition). However, the supply point itself is determined by the plant populations, which depend on pollination activity for their long-term survival. Extinction of a plant species (e.g., due to low visit quality) causes the supply point to drop to zero along the corresponding rewards axis, leading to a cascade of ecological reorganization and a new equilibrium (see below). In Plant Space, the equilibrium point in the absence of pollinators is always at the origin, since all plants require pollination services to avoid extinction. 

These three quantities (ZNGIs, impact vectors, and supply point) define the conditions for stable coexistence. Pollinator populations reach equilibrium when all the corresponding ZNGIs pass through the current environmental state. In addition, the combined impact of all pollinator species must exactly cancel the supply for the environment to remain in this state. This total impact is found by multiplying each impact vector by the corresponding population density, and then summing the results. Whenever the supply point lies within the cone formed by extending all the impact vectors backwards (Fig.~\ref{fig:reward-plant}), a set of population densities can be found with a total impact equal and opposite to the supply. Each potentially stable set of coexisting species is thus represented by an intersection of ZNGIs, and coexistence is achieved whenever the supply point falls within the corresponding coexistence cone. Our analyses assume that the pollinators' ZNGIs intersect, which reflects the fact that plant-pollinator communities exist with their many coexisting species 

\subsection*{3. Conditions for adaptive pollinator coexistence on shared rewards}
The full equilibrium of the model also requires that adaptive foraging dynamics have reached a steady state. This requirement is satisfied with additional restrictions on the parameter values, which we derive by setting the pollinator growth rate $da_j/dt = 0$ in Eq. (\ref{eq:a}) and substituting into the adaptive foraging equation (\ref{eq:adapt}). We find the following equilibrium condition:

\begin{align}
0 &= \frac{G_j \alpha_{ij}}{a_j} (c_{ij}V_{ij}^s f_{ij} - \mu_j^A).
\label{eq:adapteq}
\end{align}

The term in parentheses is what the growth rate $da_j/dt$ for animal species $j$ would be if it were a specialist on plant species $i$, with $V_{ij} = V_{ij}^s$ and $\alpha_{ij}=1$. Eq. (\ref{eq:adapteq}) requires that this term vanish at equilibrium for all plant-animal pairs $i,j$ where $\alpha_{ij}\neq 0$. Substituting in the expressions for $V_{ij}$ and $f_{ij}$ from the first section of the Methods, we find the equilibrium rewards abundance $R_i^*$:

\begin{align}
R_{i}^* = \frac{\mu_j^A}{c_{ij}\tau_{ij}b_{ij}}.
\label{eq:Rstar}
\end{align}
This result imposes a strict constraint on the animal mortality rates $\mu_j^A$ and the reward uptake efficiencies $c_{ij}\tau_{ij} b_{ij}$, requiring that both terms vary in the same way from species to species, for all animals that share rewards from the same plant species $i$ (i.e., for all animals with $\alpha_{ij} \neq 0$). \cite{mori2019adaptive} suggests that this required correlation between mortality rates and ingestion rates is consistent with allometric scaling relationships (\citealt{yodzis1992body}).

In Appendix A, we show that $R_i^*$ is also the rotation center for the ZNGI's and, therefore, the shared $R_i^*$ remains the point of intersection for all the ZNGIs over the entire course of adaptive foraging dynamics (see Appendix A). Note that the introduction of alien pollinators analyzed below does not satisfy the relationship implied by Eq. (\ref{eq:Rstar}), with the result  that the natives stop foraging on all the plants species shared with the alien.

\subsection*{4. Using projections to analyze high-dimensional ecosystems}

The graphical analysis described above is easily visualized for environmental spaces with two dimensions. Plant-pollinator networks, however, contain tens to hundreds of plant species. In this full space, the ZNGIs are no longer lines but hypersurfaces of dimension $P-1$ (Fig.~\ref{fig:nested}\emph{b}, where $P$ is the number of plant species in the network). The intersections among these hypersurfaces determine the points of potential coexistence. We extend our graphical approach to many dimensions and analyze the conditions for coexistence among the species whose ZNGI hypersurfaces intersect by using projections of the coexistence cone onto two-dimensional slices through the full environmental space.

We consider the two-dimensional slice where two of the rewards (or plant) abundances are allowed to vary (gray plane in Fig.~\ref{fig:nested}\emph{b}), while all other abundances are held fixed at the values where the intersection occurs. We then create a diagram like those of Fig.~\ref{fig:reward-plant} by drawing the lines where the ZNGIs intersect this slice, and projecting the impact vectors and supply point onto this slice (i.e., taking the component parallel to the slice's surface). The species do not coexist if the projection of the supply point lies outside the projection of the coexistence cone (e.g., Fig.~\ref{fig:nested}\emph{a-c}), because this can only happen when the actual supply point lies outside the full coexistence cone. But the supply point may still lie outside the cone (along one of the directions that has been projected out) even if the projected supply point lies inside the projected coexistence cone. To guarantee coexistence, one must examine all possible two-dimensional projections and ensure that the supply point is inside the cone in every projection (Fig. \ref{fig:projections}).

\section*{Results}

\subsection*{Effects of nestedness on network dynamics without adaptive foraging}

\begin{figure*}
	\includegraphics[width=16cm]{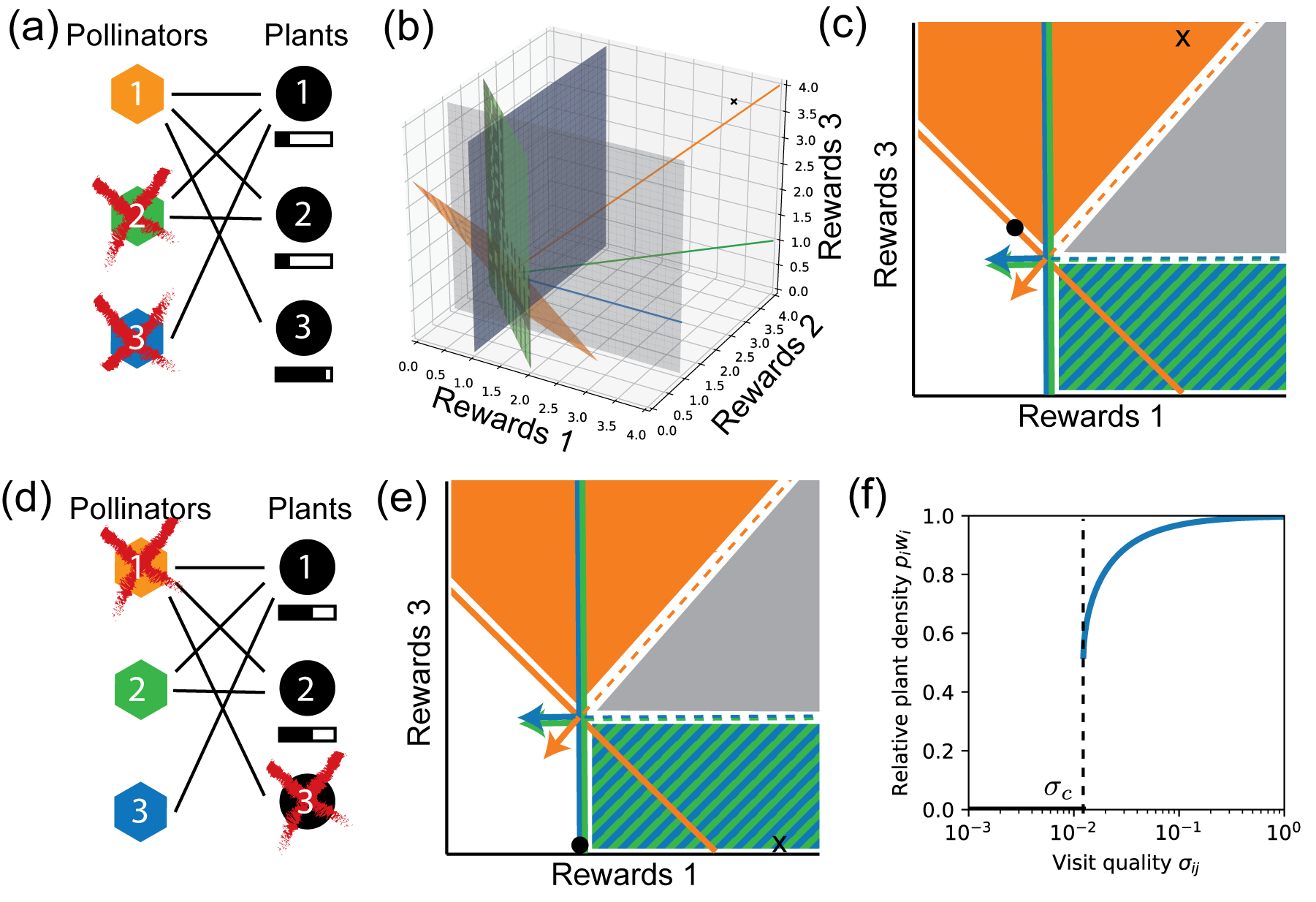}
	\caption{\linespread{1.3}\selectfont{} {\bf Effects of nestedness without adaptive foraging.} \emph{(a)} Nested network with three pollinator (polygons) and two plant (circles) species. Shaded bars indicate rewards abundance at the equilibrium point in panels \emph{c} and \emph{e}, with differences among species exaggerated for clarity. Red `x' indicates extinction at equilibrium. \emph{(b)} Three-dimensional ZNGIs, impact vectors, and supply point of this network. \emph{(c)} ZNGIs and impact vectors projected onto the the rewards 1-rewards 3 plane (gray transparent plane in \emph{b}). Pollinator species 2 and 3 have same projections onto this plane because both visit plant species 1 and none visit plant species 3 (see other projections in Fig. \ref{fig:projections}). Black dot indicates rewards at equilibrium. Specialist pollinators 2 and 3 go extinct because supply point (black `x') falls in the orange zone. \emph{(d)} Specialist plant species 3 goes extinct when the quality of visits it receives is lower than the threshold $\sigma_c$ of panel \emph{c}. \emph{(e)} Supply point drops to zero along the rewards 3 axis when plant species 3 goes extinct, which results in the extinction of the generalist pollinator species 1. \emph{(f)} Dependence of specialist plant abundance $p_i$ on visit quality $\sigma_{ij}$, using Eq.~(\ref{eq:plantpop}) from Appendix C. Minimal visit quality $\sigma_c$ required for plant persistence is indicated by the dotted line. Parameters values are taken from \cite{valdovinos2013adaptive}, with: $\tau_{ij} = 1, e_{ij} = 0.8, \mu_i^P = 0.008, c_{ij} = 0.2, \mu_j^A = 0.004, b_{ij} = 0.4, g_i = 0.4, w_i = 1.2, \beta_i = 0.2, \phi_{ij} = 0.04$. Plant abundance is measured in units of the plant's carrying capacity $1/w_i$, so that the maximum possible value equals 1. }
	\label{fig:nested}
\end{figure*}

Most plant-pollinator networks exhibit a nested structure (definition and citations provided in the Introduction). The implications of nestedness for the stability of these networks have been a topic of study for over a decade (\citealt{bastolla2009architecture,allesina2012stability}, reviewed in \citealt{valdovinos2019mutualistic}). \cite{valdovinos2016niche} provide a more mechanistic framework to evaluate the effects of nestedness on the dynamics of plant-pollinator networks. This section analytically confirms their numerical results when pollinators are fixed foragers (Eq. \ref{eq:fixed}), and provides criteria for plant survival not found by previous work (see next section for adaptive foragers).

We perform our graphical analysis using two-dimensional slices through the full rewards space of a nested 3-plant-3-pollinator-species network (Fig. \ref{fig:nested}\emph{a}), which has sufficient complexity to illustrate all the relevant projections for arbitrarily large networks. Fig. \ref{fig:nested}\emph{b} shows the three-dimensional rewards space, with the three colored planes being the ZNGIs of the three pollinator species (derived from Table \ref{tab:map}). The coexistence cone is the three-sided solid bounded by planes connecting the backwards extensions of the impact vectors (colored lines). We project this cone onto the gray transparent plane composed by rewards 1 and 3. This projection is depicted in Fig. \ref{fig:nested}\emph{c}, which shows the asymmetric shape of the coexistence cone, bounded on one side by the impact vectors of the specialist pollinators (green and blue vectors parallel to rewards axis 1), and on the other by the impact vector of the generalist pollinator species (diagonal orange vector). This asymmetric shape is characteristic of nested networks since nestedness increases the diet overlap between specialist and generalist species. This is one of only three possible cone shapes in a two-dimensional projection (see Supplementary Fig. \ref{fig:projections-all}), regardless of the full environment dimension. The second shape consists of the entire quadrant, which corresponds to the case of plant species interacting with separate subsets of pollinator species, and which always contain the supply point if the respective plant species persists. The third consists of a vanishing cone with all impact vectors pointing in the same direction, which emerges when all plant species share all pollinator species.

\cite{valdovinos2016niche} show that increasing nestedness increases the extinction of specialist species in networks without adaptive foraging. Our graphical approach explains this result by demonstrating that the asymmetric coexistence cone found most frequently in nested networks favors the extinction of specialist pollinators. To show this, we note that obtaining a supply point in the orange region of Fig. \ref{fig:nested}\emph{c} (where both specialist pollinator species go extinct) only requires that the the supply level $\beta_3 p_3/\phi_3$ of rewards 3 is greater than the supply of rewards 1. This happens 50 percent of the time when the plant parameters are randomly chosen (as they were in the previous simulations).  But for the supply point to reach the blue and green region, where one or both of the specialist pollinator species persist, the supply of rewards 3 must drop below the ZNGI intersection. This is a much more stringent condition, and in practice it is only satisfied when the specialist plant (here plant species 3) goes extinct (Fig. \ref{fig:nested}\emph{e}).

To elucidate the conditions for plant extinction in these networks, we distinguish two drivers of species elimination: competitive exclusion by other plant species for resources other than pollination, and failure to receive sufficient pollination. Plant competition is modeled with a Lotka-Volterra type competition matrix and standard techniques from coexistence theory can be employed to study this aspect (see Appendix B). We focus on the second driver by assuming intraspecific competition much stronger than interspecific competition, which effectively gives each plant species its own niche. This leaves pollination -- particularly visit quality ($\sigma_{ij}$, see Methods) -- as the sole determinant of plant survival. Specialist plants receive the lowest quality of visits in nested networks, because they are only visited by generalist pollinators that carry diluted pollen from many other species. We find the criteria for plant survival by calculating the plant population size $p_i$ as a function of the visit quality $\sigma_{ij}$ for a perfectly specialist plant (visited by only one pollinator species). We obtained an exact analytic expression for this relationship (Eq.~\ref{eq:plantpop} of Appendix C), which is depicted in Fig. \ref{fig:nested}\emph{f}. This relationship shows that each plant species remains near its maximum abundance ($1/w_i$) as long as the visit quality they receive is above a threshold $\sigma_c$, but it suddenly drops to zero when the visit quality drops below this threshold.

\subsection*{Effects of adaptive foraging}
Adaptive foraging (Eq.~\ref{eq:adapt}) rotates the ZNGIs and impact vectors in the direction of the more plentiful floral rewards (see Methods). This section explains the consequences of this rotation for species coexistence and provides analytical understanding for the result found by previous simulations showing that adaptive foraging increases the species persistence of nested networks (\citealt{valdovinos2016niche}).

\begin{figure*}
	\includegraphics[width=16cm]{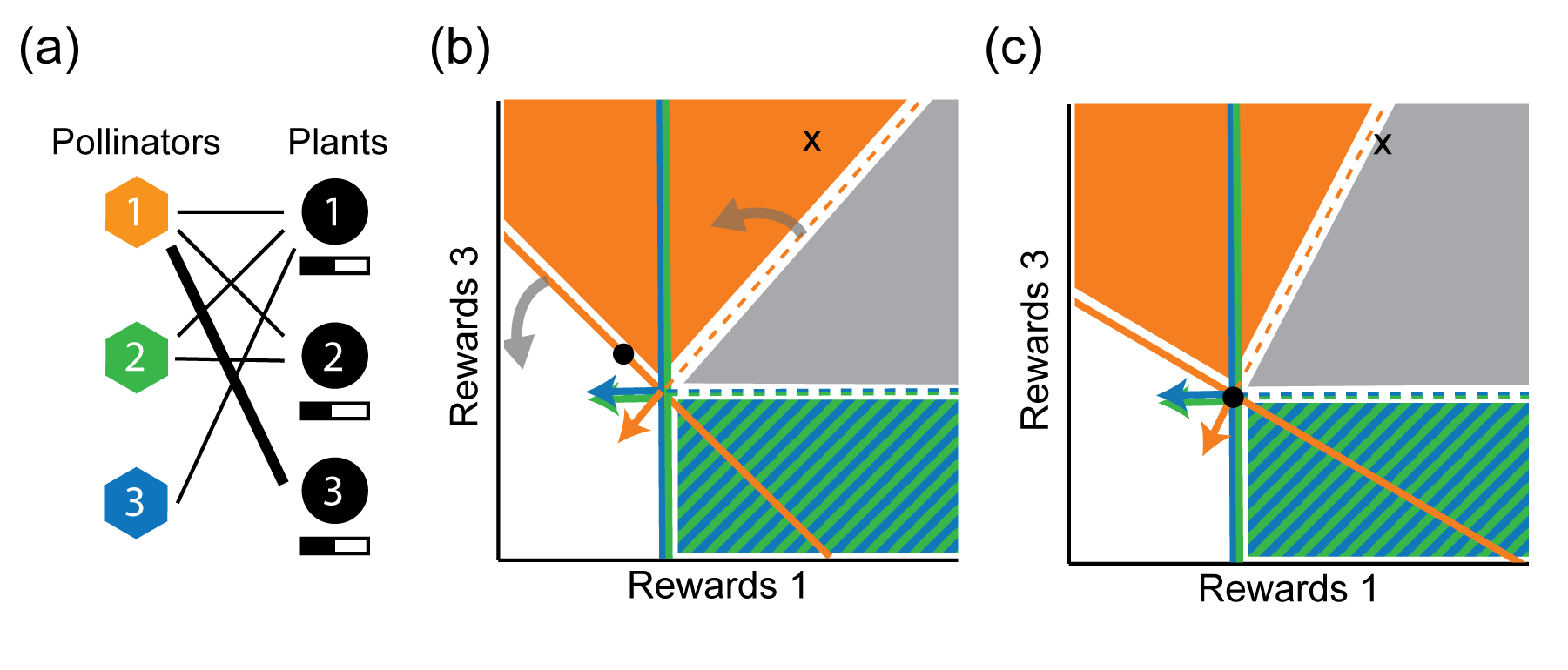}
	\caption{\linespread{1.3}\selectfont{} {\bf Effects of adaptive foraging.} \emph{(a)} Adding adaptive foraging to the nested network allows the generalist pollinators to focus their foraging effort on the plant species with more abundant floral rewards (thick line connecting pollinator 1 to plant 3). \emph{(b)} Adaptive foraging causes the ZNGI of the generalist pollinator species and its impact vector to rotate counterclockwise (towards the most plentiful rewards 3). Black dot represents the equilibrium state of Fig.~\ref{fig:nested}\emph{c}, with more available rewards in plant species 3 than in species 1.  \emph{(c)} The rotation of the impact vector expands the coexistence cone making it to engulf the supply point, so that all three species coexist in the new equilibrium (black dot). This rotation also reduces the angle between the impact vector and the rewards 3 axis, increasing the quality of visits by the generalist pollinators to these plants, while decreasing their quality of visits to the other plant species.}
	\label{fig:adaptive}
\end{figure*}

Fig. \ref{fig:adaptive} shows how adaptive foraging changes the result illustrated in Fig. \ref{fig:nested}\emph{a-c}. The supply point lies just outside the coexistence cone, and the equilibrium state with fixed foraging preferences gives plant species 3 a higher equilibrium concentration of floral rewards. This means that the generalist pollinators will begin to focus their foraging efforts on plant species 3, resulting in a rotation of the ZNGI and impact vector to become more like those pollinators specialized on plant species 3 (i.e., a horizontal line and vertical arrow in this visualization). This rotation opens up the coexistence cone until it engulfs the supply point. The resource abundances then relax to the coexistence point $(R^*_1, R^*_2, R^*_3)$, where all plants are equally good food sources, and adaptation stops. This process allows the coexistence of all pollinator species and explains how adaptive foraging increases the species persistence of pollinators.

Adaptive foraging increases coexistence among plant species in nested networks by causing pollinator species to focus their foraging efforts on more specialist plant species (Fig.~\ref{fig:adaptive}\emph{a}), increasing the visit quality they receive (see angle of the orange impact vector becoming more parallel to the rewards 3 axis in the sequence of Fig.~\ref{fig:adaptive}\emph{b-c}). This rotation in ZNGIs, in turn, decreases the visit quality that the generalist plants receive from the generalist pollinators (see angle of the orange impact vector becoming more perpendicular to the rewards 1 axis in the sequence of Fig.~\ref{fig:adaptive}\emph{b-c}). The generalist plant species will still persist despite this reduction in visit quality by generalist pollinators, because they still receive perfect visit quality from specialist pollinators that only visit them (e.g., pollinator species 3 in Fig.~\ref{fig:adaptive}\emph{a}) and which cannot shift their foraging effort to other plant species. Overly-connected networks (i.e., with many more interactions than the ones found in empirical networks) lack these perfect specialists and, therefore, the average quality of visits received by generalist plant species drops below the threshold $\sigma_c$ (Fig.~\ref{fig:nested}\emph{c}) and they go extinct, as observed in previous simulations.

\subsection*{Impact of pollinator invasions on native species}

\begin{figure*}
	\includegraphics[width=16cm]{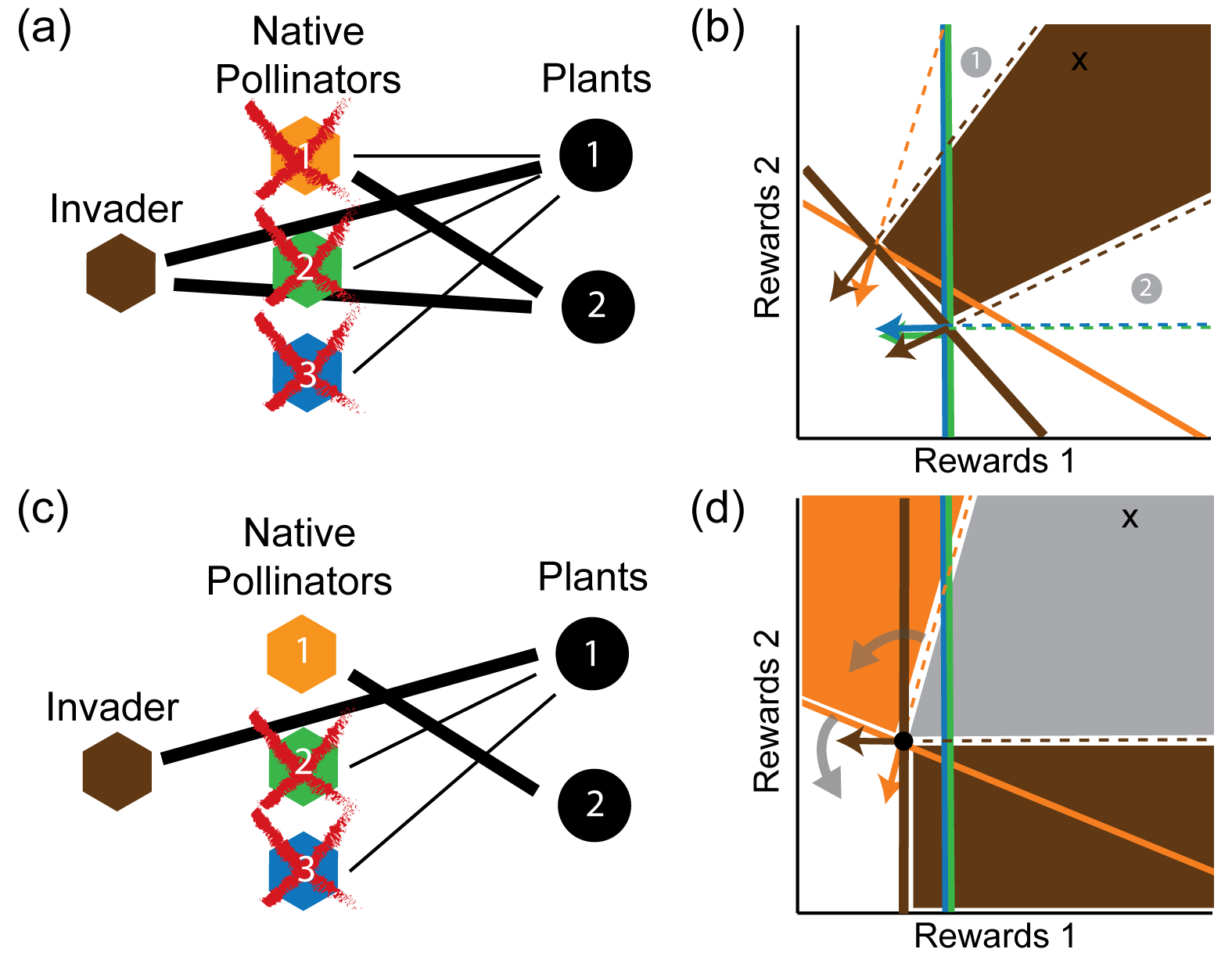}
	\caption{\linespread{1.3}\selectfont{} {\bf Pollinator invasions.} \emph{(a)} Brown polygon represents an alien pollinator species with higher visit efficiency than natives, visiting the two plant species. \emph{(b)} If plant species have similar abundances (as in previous simulations), the supply point falls in the gap between the two coexistence cones, and only the invader survives at equilibrium.  \emph{(c)} Invader does not interact with plant species 2. \emph{(d)} The supply point now falls inside the coexistence cone 1 and the invader coexist with pollinator species 1. Adaptive foraging drives the native species to become a pure specialist on plant species 2 (which had more rewards). This results in plant species 2 receiving more and better visits, and in pollinator species 1 reducing its population size. The relative abundances can be estimated from the position of the supply point within the cone. For example, only a small contribution will be required from a pollinator species to achieve perfect cancellation of the supply if one of the other impact vectors points almost directly away from the supply point. The invader's impact vector points in slightly different directions at the two coexistence points. This results from the factor of $R_i$ contained in the $f_{ij}$ term of the impact vector as given in Table \ref{tab:map}, which biases the vector in the direction of the more abundant reward. Plant extinctions do not occur under these conditions.}
	\label{fig:invasions}
\end{figure*}

This final section analyzes the consequences of pollinator invasions on species coexistence in networks with adaptive foraging, and provides analytical understanding for the results found numerically by \cite{valdovinos2018species}. We assume that alien species come from a different regional pool, with consumption and mortality rates not following the strict relationship imposed on the native species (see Methods Eq. \ref{eq:Rstar}). This results in the alien's ZNGI not passing through the natives' common ZNGI intersection (Fig. \ref{fig:invasions}\emph{b,d}), but instead intersecting different native ZNGI's at different points. The resulting proliferation of possible coexistence points and cones impede the analysis of high-dimensional systems using the method of projections employed above. Therefore, we focus on a similar network than in previous sections but with only two (instead of three) plant species.

Alien species will invade the network whenever the native coexistence point $R^*_i$ (see Methods Eq. \ref{eq:Rstar}) falls on the positive growth rate side of the alien's ZNGI, regardless of the number of plant species the alien visits. This corresponds to the case of efficient foragers reported in previous simulations (i.e., with higher foraging efficiency than natives), which were the only aliens that invaded the networks among all the alien types analyzed by \cite{valdovinos2018species}. The impact of the invader on native species will depend on how the alien's ZNGI alters the coexistence points which, in turn, depends on the network structure.

A network structure with native pollinator species visiting only plant species visited by the efficient invader (Fig. \ref{fig:invasions}\emph{a}), has three possible outcomes depending on the position of the supply point: i) native specialists go extinct when the supply point falls in the invader-generalist coexistence cone (cone 1 in Fig. \ref{fig:invasions}\emph{b}); ii) generalists go extinct when the supply point falls in the invader-specialist coexistence cone (cone 2 in Fig. \ref{fig:invasions}\emph{b}); iii) all native pollinator species go extinct when the supply point falls in the gap between the two coexistence cones (dark region in Fig. \ref{fig:invasions}\emph{b}). This third outcome (illustrated in Fig. \ref{fig:invasions}\emph{a}) happens when all plant species have similar properties (as assumed in previous simulations) which results in a supply point near the diagonal of the rewards space. This explains the result found by previous simulations of native pollinators always going extinct when they visited only plant species also visited by the efficient invader. 

A network structure where native pollinators visit plant species not visited by the invader results in the coexistence between the invader and the natives that have access to those alternative resources. For example, the pollinator species 1 coexists with the invader if the invader only interacts with plant species 1. This results in plant species 2 having higher rewards than species 1 at the new coexistence point, which makes pollinator species 1 shift its foraging effort to plant species 2 until it becomes a pure specialist (Fig. \ref{fig:invasions}\emph{c}). Conversely, all three pollinator species coexist as specialists on plant species 1 if the invader only interacts with plant species 2.

This analysis suggests that native pollinators only visiting plants visited by the invader will typically be driven extinct in larger networks, because the supply point will most likely fall in the gap between the high-dimensional coexistence cones. But if a pollinator species interacts with at least one plant species not visited by the invader, it will survive and transfer all its foraging effort to these plants. This agrees with what was observed in previous simulations.

\section*{Discussion}
%Mutualistic systems sustain the productivity and biodiversity of most ecosystems in the planet (\citealt{thompson1994coevolutionary,peay2016mutualistic,potts2016safeguarding}). Therefore, understanding the mechanisms by which their hundreds of species coexist is of major relevance for answering fundamental questions in ecology and also informing conservation efforts. Here, we advance the understanding of species coexistence in mutualistic systems by expanding Contemporary Niche Theory to analyze the dynamics of plant-pollinator networks. By applying analytical techniques from Niche Theory to the mutualistic model of \cite{valdovinos2013adaptive,valdovinos2016niche,valdovinos2018species}, we demonstrated how successfully accessing the floral rewards required by pollinators and the visit quality required by plants depend on the network structure and on adaptive foraging. We further showed how invasions by efficient pollinator species alter species coexistence, causing the extinction of native pollinators depending on the network structure.

Previous studies on species coexistence in plant-pollinator systems mainly consisted of work developing conceptual (e.g., \citealt{palmer2003competition,mitchell2009new}) and mathematical (e.g., \citealt{levin1970competition,johnson2019coexistence}) frameworks for analyzing conditions at which species can coexist, and reviews of empirical cases showing competition among plant species for pollination services (e.g., \citealt{mitchell2009new,morales2009meta}) and among pollinator species for floral rewards (e.g., \citealt{palmer2003competition}). The Contemporary Niche Theory allows a synthesis of all this information in one framework, and makes quantitative predictions about community dynamics including species coexistence. \cite{peay2016mutualistic} recently expanded this theory to represent mycorrhizal mutualisms, which allowed him to analyze the requirements for survival (requirement niche) and the impact on the environment (impact niche) of the interacting species, and to evaluate species coexistence. We further expand this theory by incorporating plant-pollinator systems. Our contributions consist of considering short- and long-term dynamics of plant-pollinator interactions, depicting the requirement and impact niches of pollinators, and demonstrating the effect of adaptive foraging and network structure on those niches. We applied these advances to the understanding of pollinator invasions. We next explain each of these contributions and contextualize them with previous literature.

\subsection*{Explicit consideration of two timescales: Rewards and Plant Spaces}
Explicit consideration of timescales has been recently highlighted as paramount for analyzing ecological systems, especially when evaluating management strategies (\citealt{callicott2002choosing,hastings2016timescales}) where the timeframe of action determines the ecological outcome. This is particularly the case of plant-pollinator systems whose dynamics can be distinctively divided into at least two timescales, the short-term dynamics occurring within a flowering season and the long-term dynamics occurring across flowering seasons. We developed our graphical approach for these short- and long-term dynamics by representing the pollinators’ niches in Rewards and Plant Spaces, respectively. Rewards Space assumes approximately constant plant populations, analyzing the dynamics occurring during a flowering season where plants do not reproduce but produce floral rewards that are depleted by pollinators in a matter of hours or days. Plant Space represents the longer timescale at which the quality and quantity of pollinator visits impact plant populations represented on the axes.

The other work we know expanding Contemporary Niche Theory to mutualisms uses a more classic consumer-resource space, where niche axes represent resources in the soil used by plant species (\citealt{peay2016mutualistic}) indistinctly of the timescale. That work shows how the plants' ZNGIs change when the mycorrhizal mutualism is added, but the axes are still resources in the soil, not mutualists. In our work, by contrast, the axes are the abundances of the mutualistic partners themselves (Plant Space) or the rewards produced by them (Rewards Space).

\subsection*{Depicting the pollinators’ requirement and impact niches}
Analysis of the requirement niches of species sharing resources has been long used to study species coexistence (\citealt{macarthur1970species,tilman_resource_1982,leibold1995niche,chase2003ecological}). Only recently has such analysis been applied to mutualistic systems. \cite{johnson2019coexistence} applied Tilman's Resource Ratio Theory to two pollinator species competing for the rewards provided by one plant species, and when an abiotic resource is added. 
%Our results in Rewards Space without adaptive foraging are similar to their results in two ways. First, both studies find that two pollinator species can coexist when competing for a plant species' rewards if they need the same amount of rewards for balancing out their mortality and reproduction rates, which we extend to any number of pollinator species with Eq.~(\ref{eq:Rstar}). Second, 
%Our work reinforces their finding that the competition outcome (pollinator coexistence, or extinction of one or both species) affects the plant population providing the rewards. Contrarily 
Our results expand this work by extending to networks with larger numbers of plant and pollinator species, where nestedness and adaptive foraging become relevant properties. However, we do not explicitly consider resources or abiotic limitations other than floral rewards that species might require to survive (e.g., nesting sites, water), which represents an important avenue for future work.

We study the pollinators' impact niche corresponding to the change induced on plant and reward abundances. In Plant Space, the mutualism is directly visible in the impacts, which represent the number of successful pollination events caused by each pollinator, and the impact vectors point in the direction of larger plant population sizes. This space shows a main difference between resource competition in classic consumer-resource and mutualistic systems. Consumers in classic consumer-resource systems can only affect each other negatively through depleting their shared resource, while consumers in mutualistic systems can also benefit each other through benefiting their shared mutualistic partner. In Rewards Space, the impact of a pollinator species is simply the rate at which it depletes the floral rewards, just as in a classic model of resource competition. An important difference, however, is the representation of the visit quality of a particular pollinator species to a particular plant species in terms of the angle between its impact vector and the rewards axis corresponding to the plant species. The analysis of this representation advances another subject that has captured the attention of ecologists for over a century, plant competition for pollination (reviewed in \citealt{mitchell2009new}). This large body of research has shown that plant species sharing the same pollinator species potentially compete not only for the pollinators’ quantity of visits but also for their quality of visits. Our approach provides means for analyzing plant competition for quantity and quality of visits quantitatively and, therefore, complements previous empirical and conceptual approaches.

Finally, the strict constraint on pollinator parameter values given by Eq.~(\ref{eq:Rstar}) highlights the intrinsic incompleteness of any model (including ours) that focuses exclusively on plant-pollinator interactions, which are only a subset of the full ecosystem (\citealt{hale2020mutualism}). Our analysis deals mainly with collections of pollinators that are already assumed to be capable of coexistence for some sets of rewards supply levels, with all ZNGIs intersecting at the point defined by Eq.~(\ref{eq:Rstar}). Questions on how many pollinator species can coexist or how to prevent competitive exclusion (\citealt{gause1935behavior,levin1970community,McGehee1977}) present interesting avenues for further study in models that consider the broader ecological and evolutionary context of plant-pollinator interactions.

\subsection*{Effects of network structure and adaptive foraging on species coexistence}
The network structure of plant-pollinator systems influences community dynamics and species coexistence by determining who interacts with whom and which mutualistic partners are shared between any two given species. We analyzed the effects of nestedness on species persistence in these networks by depicting the dynamics occurring in systems where generalist and specialist pollinators share the floral rewards of generalist plants, while specialist plants are visited only by generalist pollinators. We provided analytical understanding to results found by previous simulations by showing how nestedness with its increased niche overlap produces an asymmetric coexistence cone that causes the extinction of specialist species.

We demonstrated that adaptive foraging rotates the pollinators' ZNGIs and impact vectors towards the most abundant rewards, promoting pollinator coexistence in nested networks through niche partitioning and plant coexistence through the increased visit quality to specialist plants. We anticipate that our graphical representation of adaptive foraging can be applied to other types of consumer-resource systems such as food webs, where the effects of adaptive foraging have been extensively studied theoretically (reviewed in \citealt{valdovinos2010consequences}). For example, \cite{kondoh2003foraging} shows how adaptive foraging causes many species to coexist in complex food webs. Key to this result is the ``fluctuating short-term selection on trophic links'', which effectively reduces the realized food-web connectance. That is, adaptive foraging allows the rare prey to recover by making the consumers effectively specialize on the most abundant prey, which results in the rare prey becoming more abundant and the abundant prey becoming more rare, causing the adaptive consumers to switch their preferences again. This is similar to our result of generalist pollinators becoming effectively specialized on specialist plants with initially higher reward abundance, but is also different because our plant-pollinator model does not exhibit fluctuations in foraging preferences. This difference is explained by the inherent timescales of rewards and prey dynamics, where the rewards are produced and consumed at the same short timescale, while the production of new prey are lagged behind the consumption by predators. We anticipate that our graphical approach will deepen the conceptual unification of theory on mutualistic systems and theory on food webs, by providing analytical understanding of species coexistence in consumer-resource systems, and incorporating the effects of adaptive foraging and network structure, both critical for the dynamics of those two types of consumer-resource systems.

\subsection*{Conclusion}
Our graphical approach  promotes the unification of niche and network theories by incorporating network structure and adaptive foraging into the graphical representation of species' niches. This approach also deepens the synthesis of mutualistic and exploitative interactions within a consumer-resource framework, by including both in the graphical representation of pollinators' niches. This research may promote further development of ecological theory on mutualisms, which is crucial for answering fundamental questions and informing conservation efforts.

\section*{Acknowledgments}
We thank George Kling for his comments on earlier versions of this manuscript. This research was funded by US NSF (DEB-1834497) to FSV.

\bibliographystyle{amnatnat}
\bibliography{references.bib}

\newpage{}
\renewcommand{\theequation}{A\arabic{equation}}
%% redefine the command that creates the equation number.
\renewcommand{\thetable}{A\arabic{table}}
\renewcommand{\thefigure}{S\arabic{figure}}
\setcounter{equation}{0}  % reset counter 
\setcounter{figure}{0}
\setcounter{table}{0}

\section*{Appendix A: Analysis of adaptive foraging equation}
In this appendix, we show that the adaptive foraging dynamics given in Eq.~(\ref{eq:adapt}) of the main text cause the ZNGI of a pollinator species $j$ to rotate about a point in rewards space, whose coordinates are given by the minimum reward abundance $R_{ij}^*$ required for the pollinator to survive under a pure specialist strategy focused on plant species $i$. 

First of all, setting $da_j/dt=0$ in Eq.~(\ref{eq:a}) of the main text, with $\alpha_{ij} = 1$ and $\alpha_{kj} = 0$ for all $k \neq i$, we obtain the equilibrium condition under the pure specialist strategy:
\begin{align}
0 =c_{ij} \tau_{ij}b_{ij}a_j R_i -\mu_j^A a_j.
\end{align}
Solving for the reward abundance, we obtain:
\begin{align}
R_{ij}^* = \frac{\mu_j^A}{c_{ij}\tau_{ij}b_{ij}}.
\end{align}
This is the same as Eq. (\ref{eq:Rstar}) of the main text, but we have added an index $j$ to indicate that this point can in general be different for each pollinator species, depending on the choice of parameters.

Next, we confirm that the adaptive foraging dynamics of Eq.~(\ref{eq:adapt}) preserve the constraint $\sum_{i\in P_j} \alpha_{ij} = 1$ imposed in the initial conditions, by computing
\begin{align}
\frac{d}{dt} \sum_{i\in P_j}  \alpha_{ij} &= G_j \sum_{i\in P_j}  \alpha_{ij}\left(c_{ij}\tau_{ij} b_{ij} R_i - \sum_{k \in P_j} \alpha_{kj}c_{kj}\tau_{kj}b_{kj}R_k\right)\\
&= G_j \left(1- \sum_{i\in P_j}  \alpha_{ij}\right) \sum_{k \in P_j} \alpha_{kj}c_{kj}\tau_{kj}b_{kj}R_k.
\end{align}
Thus if $\sum_{i\in P_j} \alpha_{ij} = 1$ at any point in time, the derivative vanishes, and it remains equal to this value for all times.

Finally, we show that this constraint on the sum of $\alpha_{ij}$ guarantees that the point $R_{ij}^*$ defined above always lies on the ZNGI, i.e., that $da_j/dt$ always vanishes there:
\begin{align}
\frac{da_j}{dt} &=  \sum_{i\in P_j} c_{ij} \alpha_{ij} \tau_{ij}b_{ij}a_j R_{ij}^* -\mu_j^A a_j\\
&= \sum_{i\in P_j} \alpha_{ij} \mu_j^A a_j -\mu_j^A a_j = 0.
\end{align}

\renewcommand{\theequation}{B\arabic{equation}}
%% redefine the command that creates the equation number.
\renewcommand{\thetable}{B\arabic{table}}
\setcounter{equation}{0}  % reset counter 
\setcounter{table}{0}
\section*{Appendix B: Conditions for coexistence among plant species}
Unlike the population growth rate of pollinators that entirely depends on rewards abundances, the population growth rate of plants in the Valdovinos et al. model considers other factors (e.g., space or nutrient limitation) that are captured by a generic Lotka-Volterra type function of plant competition composed of intra- (or self-limitation) and inter-specific competition coefficients ($w_i$ and $u_l$, respectively) that affect plant recruitment rate ($\gamma_i$ in Eq.~\ref{eq:vsat}) and are independent of the mutualistic interaction with pollinators. The standard conditions for stable coexistence in Lotka-Volterra models therefore represent a necessary condition for plant coexistence. Whether a plant species actually persists at equilibrium also depends on whether it receives sufficient pollination services, which will be discussed in Appendix C below. 

To simplify our analysis, in the main text we focus on the case of low inter-specific competition (i.e., $u_l \ll w_i$), which is also the regime where all the relevant numerical simulations were performed (\citealt{valdovinos2013adaptive,valdovinos2016niche,valdovinos2018species}), so we can safely approximate $p_i^* \approx 1/w_i$ under conditions of adequate pollination. 

To go beyond this regime, and obtain necessary coexistence conditions with non-negligible interspecific competition, we must examine the stability of the fixed points of the plant dynamics given by Eq.~(\ref{eq:p}). To keep the problem tractable, we will treat $\alpha_{ij}$ as fixed parameters, and assume that $a_j$ quickly relax to the equilibrium value $a_j^*(p_k)$ corresponding to the current plant abundances. Under these assumptions, the stability of the plant equilibrium depends on the eigenvalues of the Jacobian matrix
\begin{align}
J_{ik} = \frac{\partial}{\partial p_k}\frac{dp_i}{dt} = \frac{\partial \gamma_i}{\partial p_k} \sum_{j \in A_i} e_{ij}\sigma_{ij}V_{ij} + \gamma_i \frac{\partial }{\partial p_k} \sum_{j \in A_i} e_{ij}\sigma_{ij}V_{ij} - \mu_i^P \delta_{ik},
\end{align}
evaluated at the equilibrium point $p_i^*$. If all eigenvalues have negative real parts, then the equilibrium is stable.

To further streamline the calculation, we will assume that $w_i = w$ for all $i$ and $u_l = u$ for all $l$. This allows us to state the results in terms of the relative strength of interspecific ($u$) versus intraspecific ($w$) competition. Evaluating the derivatives, we then find:
\begin{align}
J_{ik} = -\left(g_i\sum_{j \in A_i} e_{ij}\sigma_{ij}V_{ij}\right) [(w-u) \delta_{ik} + u] +\gamma_i \sum_{j \in A_i}e_{ij}\sigma_{ij}\alpha_{ij}\tau_{ij} p_i^* \frac{\partial a_j^*}{\partial p_k}+ \left(\gamma_i \sum_{j \in A_i}e_{ij}\sigma_{ij}\alpha_{ij}\tau_{ij} a_j^* - \mu_i^P \right)\delta_{ik}.
\end{align}
The final term in parentheses is equal to $d \log p_i/dt$ for $p_i > 0$, and so it must vanish whenever all the plants coexist. To determine the sign of the eigenvalues for the remaining portion, it is convenient to define the diagonal matrix $\mathbf{D}$ with components
\begin{align}
D_{ik} = \delta_{ik} g_i \sum_{j \in A_i} e_{ij}\sigma_{ij}V_{ij}
\end{align}
and a matrix $\mathbf{A}$ with components
\begin{align}
A_{ik} = \frac{\gamma_i \sum_{j \in A_i}e_{ij}\sigma_{ij}\alpha_{ij}\tau_{ij} p_i^* \frac{\partial a_j^*}{\partial p_k}}{g_i \sum_{j \in A_i} e_{ij}\sigma_{ij}V_{ij}}.
\end{align}
We can now write the Jacobian $\mathbf{J}$ in matrix notation as
\begin{align}
\mathbf{J} = - \mathbf{D}[(w-u)\mathbf{I}+\mathbf{U} - \mathbf{A}]
\label{eq:jac1}
\end{align}
where $\mathbf{I}$ is the identity matrix and $\mathbf{U}$ is a matrix with elements $U_{ij} = u$. 

In the low mortality limit $\mu_i^P \to 0$, the steady state occurs at $\gamma_i \to 0$, and so $\mathbf{A} \to 0$. In this case, the eigenvalues of $[(w-u)\mathbf{I}+\mathbf{U}]$ can be evaluated exactly, with one eigenvalue equal to
\begin{align}
\lambda^+ = w+(P-1)u
\end{align}
and the rest equal to
\begin{align}
\lambda^- = w-u.
\end{align}
For any symmetric matrix $\mathbf{M}$ with all negative eigenvalues (a so-called ``stable'' matrix), the product $\mathbf{D}\mathbf{M}$ with any diagonal matrix $\mathbf{D}$ with all positive entries also has all negative eigenvalues. This property of maintaining stability under multiplication by a positive diagonal matrix $\mathbf{D}$ is known as ``D-stability,'' and it has been proven that all sign-symmetric stable matrices are also D-stable (\citealt{hershkowitz2003positivity}). Applying this to the case at hand, we see that the eigenvalues of $\mathbf{J}$ are all negative if and only if $\lambda^- >0$. Thus we recover for arbitrary numbers of species the classic result of modern coexistence theory for two species: stable coexistence requires that intra-specific competition ($w$) is stronger than inter-specific competition ($u$) (\citealt{chesson2000mechanisms}).

To determine the impact of nonzero $\mathbf{A}$, we focus on the case where all pollinators are pure specialists, with identical parameters. Then $\mathbf{A}$ is proportional to the identity matrix:
\begin{align}
\mathbf{A} = \frac{1-[(P-1)u +w]p^*}{a^*}\frac{\partial a^*}{\partial p}\mathbf{I}
\end{align}
where $p_i^* = p^*$ and $a_j^* = a^*$ for all $i$ and $j$, since all the parameters are the same. Since the pollinators feed on the rewards produced by the plants, $\partial a^*/\partial p$ is always positive. The smallest eigenvalue of $[(w-u)\mathbf{I}+\mathbf{U}-\mathbf{A}]$ becomes
\begin{align}
\lambda^- = \tilde{w} - u
\end{align}
where the effective intra-specific competition coefficient $\tilde{w}$ is
\begin{align}
\tilde{w} = w - \frac{1-[(P-1)u +w]p^*}{a^*}\frac{\partial a^*}{\partial p}
\end{align}
which is always less than $w$. This means that the low-mortality criterion $w > u$ remains a necessary condition for coexistence. We conjecture that this remains true for arbitrary pollinator parameters and connectivity, because it there is no obvious reason why competition between different species of pollinators should selectively provide additional intra-specific feedback for the plants.

\renewcommand{\theequation}{C\arabic{equation}}
%% redefine the command that creates the equation number.
\renewcommand{\thetable}{C\arabic{table}}
\setcounter{equation}{0}  % reset counter 
\setcounter{table}{0}
\section*{Appendix C: Minimum visit quality for specialist plants}
We consider the equilibrium condition for a specialist plant of species $i$, which is visited by just one pollinator species $j$, obtained from Eq.~(\ref{eq:p}) by substituting in for $\gamma_i$ and $V_{ij}$ using the linear model described in the first section of the Methods in the main text. We set $u_l =0$, as discussed in the main text and in Appendix B, in order to obtain the minimal visit quality required for survival, under ideal conditions with no direct competition from other plant species. We find:
\begin{align}
0 &= \frac{dp_i}{dt} = g_i (1-w_i p_i)e_{ij}\sigma_{ij} \tau_{ij}\alpha_{ij} a_j - \mu_i^P.
\label{eq:planteq}
\end{align}
The pollinator population density $a_j$ can be found by solving the equilibrium condition for the rewards, obtained from Eq.~(\ref{eq:R}):
\begin{align}
0 &= \frac{dR_i}{dt} = \beta_i p_i - \phi_i R_i - b_{ij}\tau_{ij}\alpha_{ij}a_j R_i.
\end{align}
To solve this, we recall that in the equilibrium state of interest, where the adaptive foraging is also at equilibrium, the reward abundances are equal to $R^*_i$ as defined in Eq.~(\ref{eq:Rstar}) of the main text. Thus we arrive at:
\begin{align}
a_j = \frac{\beta_i p_i - \phi_i R^*_i}{b_{ij}\tau_{ij}\alpha_{ij}R^*_i}.
\end{align}
Substituting into Eq. (\ref{eq:planteq}), we have:
\begin{align}
0 &= g_i(1-w_i p_i)e_{ij}\sigma_{ij} \frac{\beta_i p_i - \phi_i R^*_i}{b_{ij}R^*_i}-\mu_i^P.
\end{align}
This is a quadratic equation in $p_i$, which can be solved to obtain:
\begin{align}
p_i = \frac{1}{w_i}\left[1-\frac{1}{2}\left(1-d_i \right)\left(1-\sqrt{1-\frac{4}{s_{ij}\sigma_{ij}(1-d_i)}}\right)\right]
\label{eq:plantpop}
\end{align}
where
\begin{align}
d_i = \frac{\phi_i R^*_i w_i}{\beta_i}
\end{align}
is the fraction of floral rewards that are lost to dilution when the plant population is at its carrying capacity $1/w_i$, and
\begin{align}
s_{ij} &= \frac{g_i e_{ij} \beta_i (1-d_i)}{w_i \mu_i^P b_{ij}R^*_i}
\end{align}
is the number of seedlings produced per plant lifetime under optimal conditions, where there are no other plant species nearby to contaminate the pollen, and the field is kept clear of all competing plants. Specifically, $g_i e_{ij}$ is the number of individual seedlings produced per pollinator visit, $(1-d_i)\beta_i /(\mu_i^P w_i)$ is the harvested rewards mass per unit area over the plant's lifetime (i.e., over the average lifetime of an individual plant in the corresponding stochastic version of this model), and $b_{ij}R^*_i$ is the rewards mass density harvested per visit.

\renewcommand{\theequation}{D\arabic{equation}}
%% redefine the command that creates the equation number.
\renewcommand{\thetable}{D\arabic{table}}
\setcounter{equation}{0}  % reset counter 
\setcounter{table}{0}
\section*{Appendix D: Saturating functional responses}
In the version of the model presented in the main text, which was employed in all the previous simulations, the pollinator growth rates are linear functions of rewards abundances. In reality, both the quantity of rewards extracted per visit $f_{ij}$ and the visit frequency $V_{ij}$ are likely to saturate at high rewards levels. All the qualitative results obtained in the main text apply to these more realistic models as well. In this Appendix, we provide mathematical expressions for these two types of saturation, along with the expressions corresponding to Eq. (\ref{eq:Rstar}) of the main text that specify the point $R_i^*$ in rewards space where adaptive foraging reaches a nontrivial steady state.

The original publication presenting the model (\citealt{valdovinos2013adaptive}) contained a discussion of saturating rewards extraction, with each pollinator capable of obtaining a finite quantity $b_{ij}^{\rm max}$ of rewards per visit, following Holling's Type II growth kinetics (\citealt{holling1959some}):
\begin{align}
f_{ij} &= b_{ij}^{\rm max}\frac{R_i}{\kappa_{ij} p_i + R_i}.\label{eq:fsat}
\end{align}
Setting $da_j/dt = 0$ and $\alpha_{kj} = \delta_{ik}$ in Eq.~\ref{eq:a} and substituting in with this formula for $f_{ij}$, we find that the equilibrium rewards level $R_{ij}^*$ for the specialist strategy satisfies:
\begin{align}
c_{ij}\tau_{ij}b_{ij}^{\rm max} = \mu_j^A \frac{\kappa_{ij} p_i + R^*_{ij}}{p_i R^*_{ij}}
\end{align}
This equation reveals a set of two sufficient conditions to give all pollinator species $j$ the same $R_{ij}^*$ (as required for adaptive foraging to admit of a steady state with all these species sharing rewards from species $i$): (i) the mass-specific rewards uptake rates $c_{ij}\tau_{ij}b_{ij}^{\rm max}$ for different $j$ must scale linearly with the mortality rates $\mu_j^A$, and (ii) $\kappa_{ij}$ must be the same for all $j$.

In addition to the finite capacity of a pollinator to extract rewards on each visit, it is reasonable to assume that there is a maximum number of visits that an animal can make per unit time. Using the same Type II kinetics, we obtain the following expression for the total visitation rate of pollinator species $j$ on plant species $i$:
\begin{align}
V_{ij} &= a_j\frac{\tau_{ij} \alpha_{ij} p_i}{1 + \sum_k \tau_{kj} \alpha_{kj} h_{kj} p_k + \sum_k \omega_{jk} a_k}.\label{eq:vsat}
\end{align}
Here $h_{kj}$ is the handling time for pollinator species $j$ foraging on plant species $k$, and $\omega_{jk}$ quantifies the magnitude of direct interference between pollinators. Direct interference significantly complicates the geometric interpretation, so we will set $\omega_{jk}=0$ here. If the saturation of visit frequency is the only relevant nonlinearity, and the rewards uptake per visit is still linear in $R_i$, then the ZNGIs remain linear. When both kinds of saturation are present, the specialist equilibrium point $R^*_{ij}$ is defined by:
\begin{align}
c_{ij}\tau_{ij}b_{ij}^{\rm max} = \mu_j^A \frac{(\kappa_{ij} p_i + R^*_{ij})(1 + \sum_k \tau_{kj} \alpha_{kj} h_{kj} p_k)}{p_i R^*_{ij}}.
\end{align}
Giving all species the same set of $R_{ij}^*$ requires two more assumptions beyond what was required for saturating rewards extraction alone: (i) the handling time $h_{kj}$ must be inversely proportional to the visitation efficiency $\tau_{kj}$ for all pollinator species $j$ visiting a given plant species $k$, and (ii) all the plant population densities (for non-extinct plants) must be the same. Both of these are trivially satisfied under conditions similar to the simulations discussed in the main text, where the only differences between species come from the topology of the interaction network, and all other parameters are species-independent.  

\begin{figure*}
	\includegraphics[width=16cm]{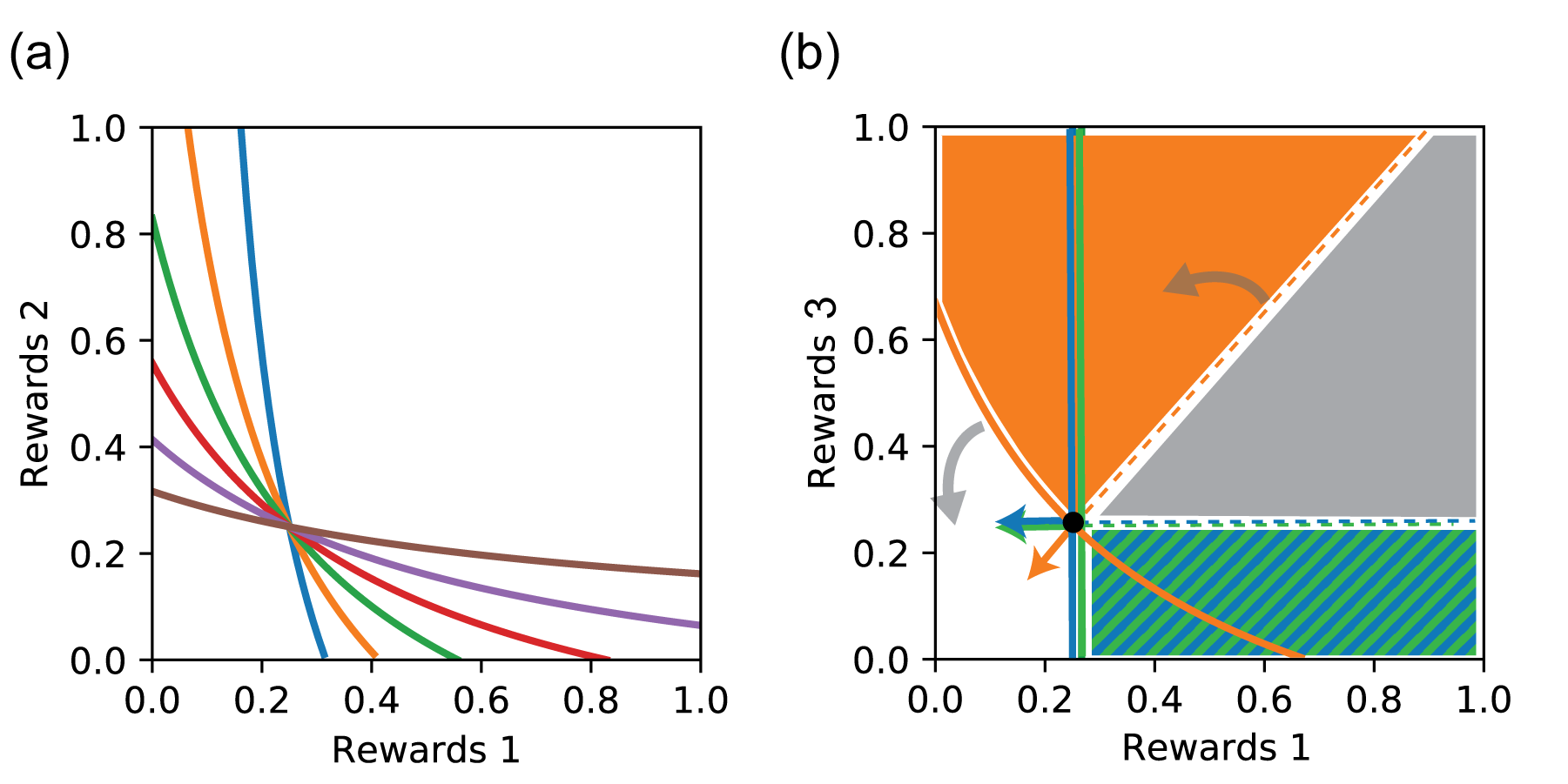}
	\caption{\linespread{1.3}\selectfont{} {\bf Saturating growth laws.} \emph{(a)} Scaling the maximum mass-specific rewards uptake rate $c_{ij}\tau_{ij}b_{ij}^{\rm max}$ with the pollinator mortality rate $\mu_j^A$ ensures that all species have the same minimum viable rewards level $R_{ij}^*$ under a specialist strategy on each plant species $i$. As in the linear model, this implies that all ZNGIs cross at this point, and rotate about it during adaptive foraging. \emph{(b)} ZNGIs, impact vectors, supply vector and coexistence cone for the nested network of Fig.~\ref{fig:nested}, with saturating rewards uptake following Eq.~(\ref{eq:fsat}). Gray arrows indicate the direction of rotation of the ZNGI and coexistence cone boundary under adaptive foraging.}
	\label{fig:saturation}
\end{figure*}

Fig.~\ref{fig:saturation} shows that the ZNGIs are no longer linear under saturating rewards extraction, but that the graphical arguments from the main text still hold. The key point is that when all parameters are species-independent (except for interaction network topology) the initial impact vectors are required by symmetry to be perpendicular to the ZNGIs, and adaptive foraging tends to rotate them away from the rewards axes corresponding to generalist plants, just as in the linear model. Since these are the two essential features necessary for recovering the simulation results, we expect that the same phenomena will be observed even in the presence of saturation.

\begin{figure*}
	\includegraphics[width=16cm]{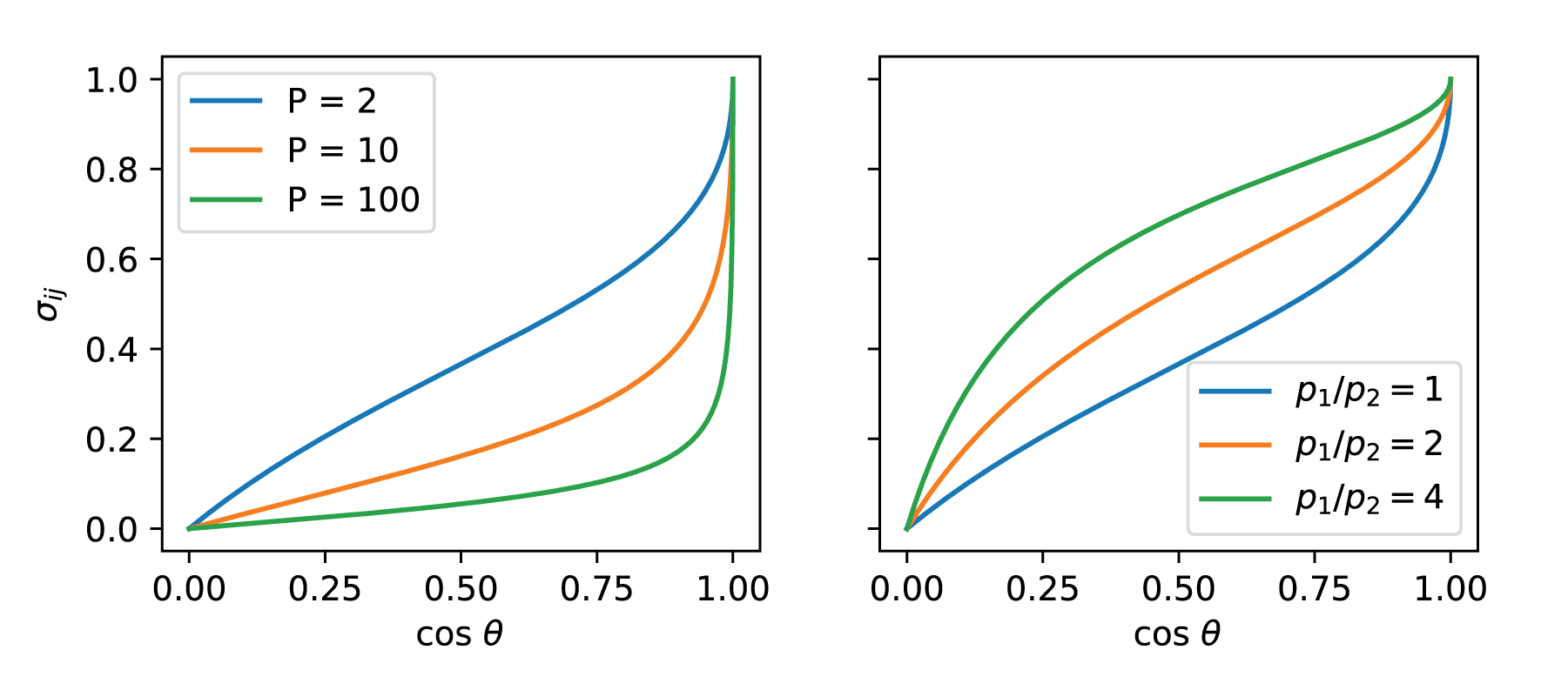}
	\caption{\linespread{1.3}\selectfont{} {\bf Relation between angle and quality.} Left: Visit quality $\sigma_{ij} = \alpha_{ij}\tau_{ij}p_i/\sum_{k\in P_j} \alpha_{kj}\tau_{kj} p_k$ versus cosine of the angle $\theta$ between the impact vector of pollinator species $j$ and (negative) rewards axis $i$ ($\cos\theta = \alpha_{ij} \tau_{ij}b_{ij}/\sum_{k\in P_j}(\alpha_{kj} \tau_{kj}b_{kj})^2$). All plants are assumed to have identical abundances $p_i$, all foraging efficiencies $\tau_{ij}$ and per-visit rewards extraction $b_{ij}$ are equal, and the foraging effort not expended on plant $i$ is equally distributed over all other plant species. Each line represents a different value of the total number of plant species $P$. Right: Same as previous panel, but for $P=2$, and different values of the ratio $p_1/p_2$ of the two plant abundances. Note that $\sigma_{ij}=0$ always corresponds to $\cos\theta=0$, and $\sigma_{ij}=1$ always corresponds to $\cos\theta = 1$, and that between these two extremes the relationship is always monotonic.}
	\label{fig:angle}
\end{figure*}

\begin{figure*}
	\includegraphics[width=16cm]{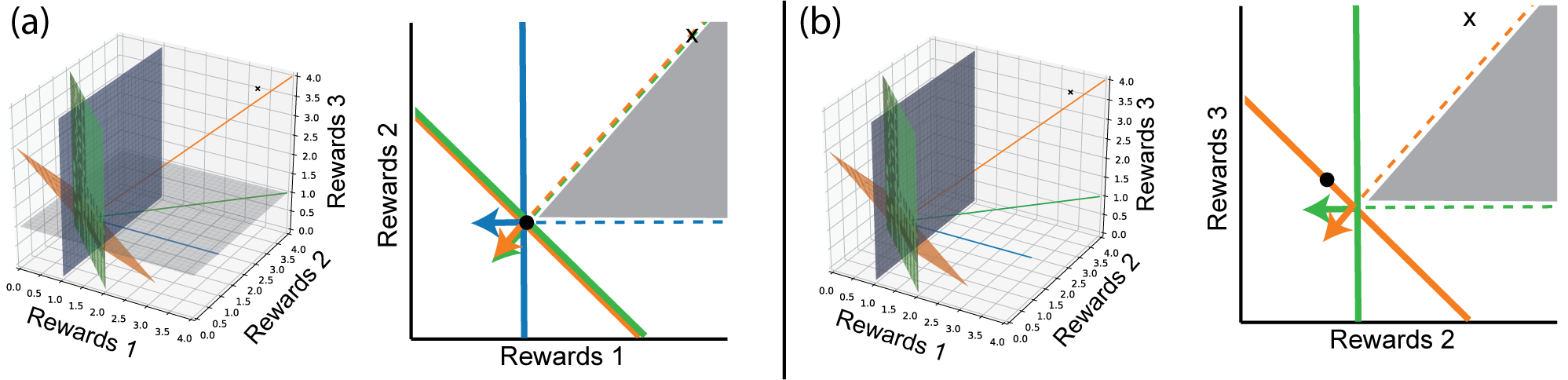}
	\caption{\linespread{1.3}\selectfont{} {\bf Additional projections.} Projections of the three-plant, three-pollinator system of Fig. \ref{fig:nested} onto the other two planes: \emph{(a)} Rewards 1/Rewards 2 \emph{(b)} Rewards 2/Rewards 3. Note that the blue species is not visible in the second projection, because the ZNGI is parallel to the projection plane, and the impact vector is perpendicular to the plane.}
	\label{fig:projections}
\end{figure*}

\begin{figure*}
	\includegraphics[width=16cm]{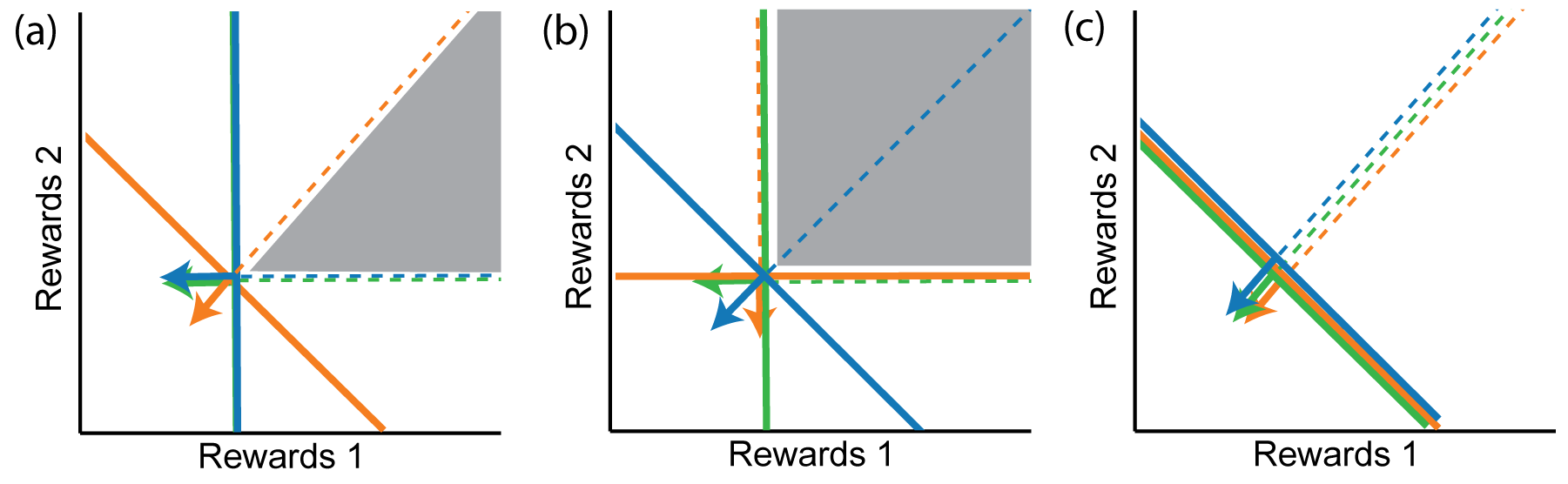}
	\caption{\linespread{1.3}\selectfont{} {\bf Complete set of possible projections without AF.} There are only three distinct two-dimensional projections of the coexistence cone that are possible in the absence of adaptive foraging. The shape of the projected cone depends only on the existence of pollinators that service one of the two plants in the projection but not the other. \emph{(a)} One plant has a specialist pollinator. \emph{(b)} Both plants have specialist pollinators. \emph{(c)} Neither plant has a specialist pollinator.}
	\label{fig:projections-all}
\end{figure*}

\end{document}